\documentclass[10pt,journal,epsfig]{IEEEtran}

\usepackage{bm}

\usepackage[dvips]{graphicx}

\usepackage{amssymb}
\usepackage{caption}

\usepackage{amsmath}
\usepackage{algorithm}
\usepackage{algorithmic}
\usepackage{multirow}
\usepackage{amsthm}
\usepackage{color}
\usepackage{epstopdf}
\usepackage{comment}
\usepackage{graphicx,subfig}
\usepackage{algorithm}
\usepackage{algorithmic}

\usepackage{cite}
\makeatletter
\let\NAT@parse\undefined
\makeatother
\usepackage[colorlinks = true,linkcolor = blue, urlcolor = blue,citecolor = blue,
anchorcolor = blue]{hyperref}
\def\BibTeX{{\rm B\kern-.05em{\sc i\kern-.025em b}\kern-.08em
		T\kern-.1667em\lower.7ex\hbox{E}\kern-.125emX}}
\begin{document}

\title{Compressive Near-Field Wideband Channel Estimation for THz Extremely Large-scale MIMO Systems}
\author{Jionghui Wang, Hongwei Wang, Jun Fang, Lingxiang Li, and Zhi Chen
\thanks{Jionghui Wang, Hongwei Wang, Jun Fang, Lingxiang Li, and Zhi Chen 
	are with the National Key Laboratory of Wireless Communications, 
	University of Electronic Science and Technology of China, Chengdu 611731, China,
	Email: hongwei\_wang@uestc.edu.cn}
\thanks{This work was supported by the Fundamental Research Funds for the Central University under Grant ZYGX2022J029 and by the Natural Science Foundation of Sichuan Province under Grant 2025ZNSFSC0514.}}

\maketitle

\begin{abstract}

 We consider the channel acquisition problem for a wideband terahertz (THz) communication system, where an extremely large-scale array is deployed to mitigate severe path attenuation. In channel modeling, we account for both the near-field spherical wavefront and the wideband beam-splitting phenomena, resulting in a wideband near-field channel. We propose a frequency-independent orthogonal dictionary that generalizes the standard discrete Fourier transform (DFT) matrix by introducing an additional parameter to capture the near-field property. This dictionary enables the wideband near-field channel to be efficiently represented with a two-dimensional (2D) block-sparse structure. Leveraging this specific sparse structure, the wideband near-field channel estimation problem can be effectively addressed within a customized compressive sensing framework. Numerical results demonstrate the significant advantages of our proposed 2D block-sparsity-aware method over conventional polar-domain-based approaches for near-field wideband channel estimation.
\end{abstract}



\section{Introduction}

The next generation of communication systems envisions extremely large-scale multiple-input multiple-output (XL-MIMO) technology and the use of the terahertz (THz) frequency band as key enabling technologies to meet the growing demands for higher data rates, ultra-low latency, and massive connectivity~\cite{zhang20236g,rappaport2019wireless}. As the aperture of the array and the frequency band increase, the Rayleigh distance can reach tens or even hundreds of meters, which is comparable to the coverage radius of the base station (BS). This indicates that most users are very likely to be located in the near-field region of the BS. Consequently, this poses a challenge for obtaining accurate channel state information, which is crucial for unleashing the potential performance of THz XL-MIMO systems.

The widely used planar wavefront assumption fails to accurately represent the propagation property of electromagnetic waves in the near-field region. Therefore, the spherical wavefront assumption should be adopted to model the channels between the BS and the users located in the BS's near-field region.
Consequently, the angular sparsity does not hold for a near-field channel. Therefore, directly applying the existing far-field channel estimation methods~\cite{lee2016channel,huang2018iterative} to estimate the near-field channel will lead to a significant performance loss. To address this challenge, a polar-domain dictionary matrix is developed in~\cite{cui2022channel} such that the near-field channel can be sparsely represented. By leveraging this specific sparse structure, the near-field channel estimation can be formulated as a sparse recovery problem, which can be efficiently solved by several well-developed compressive sensing algorithms, e.g., the orthogonal matching pursuit (OMP)~\cite{cui2022channel} and the sparse Bayesian learning methods~\cite{cao2023efficient}.

Although the utilization of the polar-domain dictionary effectively addresses the near-field channel estimation and achieves promising results, this approach may possess its inherent limitations. The polar-domain dictionary comprises a large number of atoms sampled over the angular-distance domain, often exceeding the number of the antennas by several times. This results in prohibitively high computational complexity. In addition, the atoms within the polar-domain dictionary are not mutually orthogonal, and the correlations among these atoms weaken the restricted isometry property (RIP) of the polar-domain dictionary matrix, which can adversely affect the performance of resulting near-field channel estimation algorithms. To address these challenges, a distance-parameterized approach is proposed to model the near-field channel~\cite{zhang2023near}. However, this approach requires joint estimation of the dictionary and the channel, which further increases computational complexity. In our previous works~\cite{wang2023compressive,wang2024near}, we propose a novel modified discrete Fourier transform (DFT) matrix as the dictionary. This allows the near-field channel to be represented in a block-sparse manner. Such a dictionary is essentially an unitary matrix, and hence, compared with the polar-domain dictionary, it is more amiable for compressive sensing.

On the other hand, the THz band offers abundant and unexplored spectrum
resources. Therefore, orthogonal frequency divesion multiplexing (OFDM) technology is commonly integrated to enhance the spectral efficiency. In narrowband systems with limited bandwidth, the channel across different frequencies can approximately share the same sparse structure under the same dictionary. This allows the multi-subcarrier channel estimation to be converted to a multiple measurement vector (MMV) problem, which can be solved by various methods, e.g., simultaneously OMP (SOMP)~\cite{tsai2018efficient,cui2022channel} and iterative reweighted
algorithm~\cite{shi2023channel}. In contrast, the wideband channels across different frequencies may exhibit frequency-dependent sparse structures, a phenomenon referred to as the ``beam-squint'' or ``beam-splitting'' effect.
Although extensive research has been devoted to far-field wideband channel estimation considering the beam-squint effect, these methods are inapplicable to near-field wideband channels due to the distinct differences between near-field and far-field channels. Several studies have also explored near-field wideband channel estimation. For instance, the authors in~\cite{cui2023near} revealed the bilinear sparsity of the beam-split effects in the polar domain and developed a bilinear pattern detection (BPD) based method to jointly recover channels across sub-carriers. In addition, the BPD based method was improved and extended in~\cite{lu2024subarray,wang2024mbpd}. Despite these efforts, these approaches, which are based on the polar-domain sparsity, are still subject to the limitations associated with the polar-domain dictionary.

In this paper, we investigate the near-field wideband channel estimation
problem by employing a modified DFT dictionary as a common representation dictionary for the channels across sub-carriers. It is found that the wideband channel can be sparsely represented with a two-dimensional (2D) block-sparse structure. By leveraging this specific 2D block sparsity, the wideband channel estimation problem is transformed into a 2D pattern-coupled sparse signal recovery problem, which can be efficiently addressed using the 2D pattern-coupled sparse Bayesian learning (PCSBL) algorithm. Numerical results demonstrate that, compared with existing near-field wideband channel estimation methods based on the polar-domain sparse structure, the proposed approach achieves significantly improved channel estimation accuracy.

\section{System Model and Wideband Near-Field Channel Model}

\subsection{System model}

Consider the downlink transmission of a THz XL-MIMO system using orthogonal frequency-division multiplexing (OFDM). A base station (BS) equipped with an extremely large-scale uniform linear array of $N$ antennas is deployed to serve the single-antenna users. We assume that the BS has a hybrid analog and digital precoding structure and equips $R \ll N$ radio-frequency (RF) chains.
The carrier frequency is denoted as $f_c$, and
the wavelength is $\lambda_c = {c}/{f_c}$, where $c$ is the speed of light. The antenna spacing between two adjacent antennas is set to
$d = {\lambda_c}/{2}$, and thus the antenna aperture is given by
$D = (N-1)d$. The considered system operates over a total bandwidth
of $B$ and consists of $P$ sub-carriers. The frequency associated with the $p$th sub-carrier is expressed as $f_p = f_c + \frac{2p-P}{2P}B$, with the
corresponding wavelength given by $\lambda_p =  {c}/{f_p}$.

During the downlink transmission process, the BS first sends pilot sequences to users for channel estimation. At this stage, each user individually receives the transmitted signal and estimates its own channel. Therefore, we only focus on the channel estimation of a specific user. At the $t$th time instant, the transmitted signal of the $p$th sub-carrier can be expressed as
\begin{align}
	\boldsymbol{s}_p(t) = \boldsymbol{f}(t)x_p(t),
\end{align}
where $\boldsymbol{f}(t)$ is the hybrid beamforming vector of the $p$th sub-carrier, and $x_p(t)$ is the transmitted symbol. For simplicity, we let $x_p(t)=1,\forall p,t$ during the channel estimation stage.

Denote $\boldsymbol{h}_p \in \mathbb{C}^{N \times 1}$ as the channel
from the BS to the user at the $p$th sub-carrier, and $\boldsymbol{F}
\triangleq \left[\boldsymbol{f}(1) \ \cdots \ \boldsymbol{f}(T)
\right]^H \in \mathbb{C}^{T\times N}$ represents the beamforming matrix
collecting the beamforming of all $T$ time instants. The transmitted signal received by the user at the $p$th sub-carrier is then given by \begin{align}
	\boldsymbol{y}_p = \boldsymbol{F}\boldsymbol{h}_p + \boldsymbol{n}_p,
	\label{ry_1}
\end{align}
where $\boldsymbol{n}_p(t)\in \mathcal{CN}(0,\sigma_n^2\boldsymbol{I})$
is the addictive white Gaussian noise. By stacking all the received signals in a matrix, we can obtain
\begin{align}
	\boldsymbol{Y} = \boldsymbol{F} \boldsymbol{H} + \boldsymbol{N},
	\label{ry_2}
\end{align}
where $\boldsymbol{Y} \triangleq \left[\boldsymbol{y}_1 \ \cdots \
\boldsymbol{y}_P\right] \in \mathbb{C}^{T\times P}$, $\boldsymbol{H}
\triangleq \left[\boldsymbol{h}_1 \ \cdots \boldsymbol{h}_{P}\right]
\in \mathbb{C}^{N \times P}$ and $\boldsymbol{N} \triangleq \left[
\boldsymbol{n}_1 \ \cdots \ \boldsymbol{n}_P\right]$ denote the
received signal, the wideband channel and the noise over all $P$
sub-carriers, respectively.

\subsection{Wideband Near-Field Channel Model}

The channel between BS and the user at the $p$th sub-carrier can be
represented by a geometric channel model, which is given by
\begin{align}
	\boldsymbol{h}_{p} &= \sqrt{\frac{N}{L}}\sum_{l=0}^{L-1}g_{l,p}e^{-j2\pi f_p \tau_{l}} \boldsymbol{a}(r_{l},\theta_l,f_p)
	\nonumber\\
	&= \sqrt{\frac{N}{L}}\sum_{l=0}^{L-1}\tilde{g}_{l,p} \boldsymbol{a}(r_{l},\theta_l,f_p),
	\label{hp}
\end{align}
where $\tilde{g}_{l,p}\triangleq g_{l,p}e^{-j2\pi f_p \tau_{l}}$; $L$ is the number of paths; $\{g_{l,p},\theta_l,\tau_{l},r_l\}$ are the channel gain, the angle-of-departure (AoD), the time delay, and the distance from the reference antenna of the BS to the user/scattering point associated with the $l$th path; and $\bm a(r,\theta,f_p)\in\mathbb{C}^{N\times 1}$ denotes the near-field steering vector of the ULA associated with the frequency $f_p$. Specifically, $\bm a(r,\theta,f_p)$ is given by
\begin{align}
\boldsymbol{a}(r,\theta,f_p) \triangleq \frac{1}{\sqrt{N}}\left[
		e^{-j\frac{2\pi}{\lambda_p}(r^{(1)}-r)} \ \cdots \
		e^{-j\frac{2\pi}{\lambda_p}(r^{(N)}-r)}
	\right]^T,
	\label{str_vec}
\end{align}
where $r^{(n)}$ denotes the distance between the $n$th antenna and the scatterer/user. Apparently, $r^{(1)}\equiv r$ if the first antenna is regarded as the reference antenna. According to the geometric relations, $r^{(n)}$ is given by
\begin{align}
r^{(n)} = \sqrt{r^2 + (n-1)^2d^2-2r(n-1)d\sin(\theta)}.
	\label{r_relation}
\end{align}
The expression of $r^{(n)}$ in~\eqref{r_relation} has an intractable
form for subsequent analysis. To address this issue, by applying
the second-order Taylor expansion, we can obtain the following
widely adopted approximation of  $r^{(n)}$~\cite{cui2022channel}, i.e.
\begin{align}
	r^{(n)}\approx r - (n-1)d\sin(\theta)+\frac{(n-1)^2d^2
		\cos^2(\theta)}{2r}.
	\label{r_approx}
\end{align}

Substituting~\eqref{r_approx} into~\eqref{str_vec} yields
\begin{align}
	\boldsymbol{a}(r ,\theta ,f_p) \approx \boldsymbol{a}(\theta ,f_p) \circ
	\boldsymbol{b}(\mu(r ,\theta ),f_p),
	\label{str_approx}
\end{align}
where $\circ$ represents the Hadamard product; $\boldsymbol{a}(\theta ,f_p)$ is the far-field steering vector at the frequency $f_p$, which is given as
\begin{align}
	\boldsymbol{a}(\theta ,f_p) \triangleq\frac{1}{\sqrt{N}}\left[
		 1, \cdots, e^{-j\frac{2\pi}{\lambda_p}(N-1)d\sin(\theta )}
	\right]^T,
	\label{str_decom1}
\end{align}
and $\boldsymbol{b}(\mu(r ,\theta ),f_p)$ is an unit-modulus vector given
by
\begin{align}
	\boldsymbol{b}(\mu(r ,\theta ),f_p) \triangleq \left[
		1 \ \cdots \
		e^{-j\frac{2\pi}{\lambda_p}\frac{(N-1)^2d^2}{2\mu(r ,\theta )}}
	\right]^T,
	\label{str_decom2}
\end{align}
in which $\mu(r ,\theta ) \triangleq  {r }/{\cos^2(\theta )}$,
referred to as the ``\emph{effective distance}'', is a parameter
determined by both $r $ and $\theta $. The approximation in~\eqref{r_approx} holds valid when the distance between the BS and the
user/scatterer exceeds the Fresnel distance, which is defined as $F\triangleq 0.5\sqrt{{D^3}/{\lambda}}$~\cite{wang2024near}. In practical systems, the Fresnel distance is typically orders of magnitude smaller than the coverage radius, rendering the approximation error negligible in most scenarios. Without loss of generality, we assume~\eqref{r_approx} holds strictly to simply our analysis and representation. Any residual approximation error can be considered as additive noise which can be merged into the noise term in the channel estimation problem.

There are two aspects should be noticed in the channel model shown in~\eqref{hp}. First of all,~\eqref{hp} takes the near-field effect into account. It is known that the integration
of XL-MIMO and THz band leads to a greatly expanded Rayleigh distance. In this case, a
large part of cell is covered by the near-field region of the BS. Therefore, the conventional planar wavefront model should be replaced by the spherical wavefront model, which makes the near-field steering vector depends not only on the angle but also the distance~\cite{sherman1962properties}. In addition,~\eqref{hp} is a frequency-dependent channel model. With the growth of bandwidth
and array aperture, the operating frequencies in the steering vectors cannot be approximated to
the center carrier frequency, i.e., $f_p \neq f_c$, leading to the emergence of frequency-dependent steering vectors~\cite{wang2018spatial}. This phenomenon is also known as the wideband ``beam-splitting`` effect.

\subsection{Problem Formulation}

Our objective is to extract the wideband channel $\boldsymbol{H}$ from noisy observations $\boldsymbol{Y}$. The simplest approach is the least squares (LS) method. However, the LS method has to solve an over-determined linear system to obtain a reasonable estimate, entailing a substantial amount of pilot overhead. By taking advantage of the sparse scattering characteristics of THz channels, the single-carrier far-field channel admits the angular sparsity. As a result, channel estimation can be formulated as a sparse signal recovery problem, which significantly reduces the training overhead.

However, the combined impact of the spherical effect and the beam-split effect makes wideband near-field channel estimation much more challenging. The spherical effect destroys the angular sparsity and the relevant far-field channel estimation methods~\cite{lee2016channel,huang2018iterative} suffer from severe performance degradation. The beam-splitting effect implies that if a common frequency-independent dictionary is used~\cite{wang2018spatial}, the channels at different frequencies will split into different sparse structures. Although there have been several solutions~\cite{cui2023near,lu2024subarray,elbir2023nba,wang2024mbpd} to the near-field wideband channel estimation, they are all based on the polar-domain dictionary, which has the unfavorable large dimension and the weak RIP property.

\section{Proposed Method}

\subsection{Block Sparsity of the Near-Field Channel for a Specific sub-carrier}
In the previous work \cite{wang2023compressive,wang2024near}, the numerical
results and theoretical analysis show that a single-carrier near-field
channel exhibits a block-sparse representation on a specially designed
unitary matrix. Specifically, the unitary matrix is given by
\begin{align}
		\boldsymbol{D}_{\mu} = \text{diag}(\boldsymbol{b}(\mu,f_c))\boldsymbol{D}
\end{align}
where $\mu$ is a pre-specified effective distance and $\boldsymbol{D}
\in \mathbb{C}^{N \times N}$ is a DFT
matrix with its $n$th column given by $\boldsymbol{a}(\theta_n,f_c)$
with $\sin(\theta_n) = \frac{2n-1-N}{N}, n = 1,\cdots, N$. Based on
this dictionary, $\boldsymbol{h}$ can be block-sparsely represented as
\begin{align}
\boldsymbol{h} = \boldsymbol{D}_{\mu} \boldsymbol{\beta}_{\mu}
\end{align}
where $\boldsymbol{\beta}_{\mu}$ is a block-sparse vector, which is
given by
\begin{align}
\boldsymbol{\beta}_{\mu} = \boldsymbol{D}_{\mu}^H \boldsymbol{h}
=\sum_{l=1}^{L}\tilde{g}_l \boldsymbol{D}_{\mu}^H
\boldsymbol{a}(r_l,\theta_l,f_c)
\label{sparse_sc}
\end{align}
with $\tilde{g}_{l}\triangleq g_le^{-j2\pi f_c \tau_l}$. In addition,
$\|\boldsymbol{\beta}_{\mu}\|_0$ is on the order of $1/\sqrt{N}$.

By utilizing the results in \cite{wang2023compressive,wang2024near},
given an unitary matrix $\boldsymbol{D}_{p} \triangleq
\boldsymbol{D}(\mu,f_p)$, the channel of the $p$th sub-carrier can be
mapped into a block-sparse vector, i.e.
\begin{align}
	\boldsymbol{h}_p = \boldsymbol{D}_{p}\boldsymbol{\beta}_{p}
\end{align}
Note that $\boldsymbol{D}_p$ depends on $f_p$. In other words,
$\boldsymbol{D}_p$ creates a beamspace dictionary operating at the
frequency $f_p$, in which the near-field steering vector of the same
frequency exhibits a block-sparse pattern that is related to its
physical locations $\{r_l,\theta_l\}$.

By leveraging the block-sparsity of $\boldsymbol{h}_p$, each subcarrier channel can be obtained by using several existing block-sparse recovery algorithms, e.g. the block-sparse OMP~\cite{eldar2010block} and the pattern-coupled sparse Bayesian learning (PCSBL) algorithm~\cite{duan2015pattern}. Consequently, the wideband channel can be acquired by independently employing the block-sparsity-aware algorithm with a frequency-dependent dictionary for each sub-carrier. However, this subcarrier-to-subcarrier scheme ignores the intrinsic
correlation between the channels of different sub-carriers, resulting in substantial training overhead and high computational complexity. To avoid this issue, we aim to jointly estimate the wideband channel for all
sub-carriers. Therefore, it is crucial to exploit the inherent sparse structure of $\boldsymbol{H}$.

\subsection{2D Block Sparsity of the Wideband Near-Field Channel}
To explore the joint sparsity of the wideband channel, we first construct an unified transform dictionary $\boldsymbol{D}_{\mu} \triangleq \text{diag}(\boldsymbol{b}(\mu,f_c))\boldsymbol{D}$ for all sub-carriers. This dictionary is parameterized by the center frequency $f_c$, while the channels are operating at different sub-carrier frequencies. Therefore, the channels across different frequencies will be transformed into different sparse structures on this dictionary. Specifically, the channel of the
$p$th sub-carrier can be written as
\begin{align}
	\boldsymbol{h}_p = \boldsymbol{D}_{\mu} \tilde{\boldsymbol{\beta}}_p,
\end{align}
where
\begin{align}
	\tilde{\boldsymbol{\beta}}_p = \boldsymbol{D}_{\mu}^H\boldsymbol{h}_p
= \sum_{l=1}^{L}\tilde{g}_{l,p}
\boldsymbol{D}_{\mu}^H\boldsymbol{a}(r_l,\theta_l,f_p).
	\label{sparse_fc}
\end{align}
In the following, we will show that
$\tilde{\boldsymbol{\beta}}_p$ is also a block-sparse vector and
$\|\tilde{\boldsymbol{\beta}}_p\|_0$ is also on the order of $1/\sqrt{N}$.

Recall the approximation in~\eqref{str_approx}, the near-field steering vector $\bm a(r_l,\theta_l,f_p)$ can be decomposed as the Hadamard product of a far-field steering vector
$\boldsymbol{a}(\theta_l ,f_p)$ and an unit-modulus vector $\boldsymbol{b}
(\mu(r_l ,\theta_l ),f_p)$, both of which are operating at $f_p$. In addition, $\boldsymbol{a}(\theta_{l},f_p)$ and $\boldsymbol{b}(\mu(r_l,\theta_l),f_p)$
can be equivalently expressed as
\begin{align}
	\boldsymbol{a}(\theta_l,f_p) &=\frac{1}{\sqrt{N}}\begin{bmatrix}
		&1 &\cdots
		&e^{-j\frac{2\pi}{\lambda_c}(N-1)d\eta_p\sin(\theta_l)}
	\end{bmatrix}^T
	\nonumber\\
	&=\frac{1}{\sqrt{N}}\begin{bmatrix}
		&1 &\cdots
		&e^{-j\frac{2\pi}{\lambda_c}(N-1)d\sin(\tilde{\theta}_{l,p})}
	\end{bmatrix}^T
	\nonumber\\
	&= \boldsymbol{a}(\tilde{\theta}_{l,p},f_c),
	\label{str_decom1_2}
\end{align}
and
\begin{align}
	\boldsymbol{b}(\mu(r_l,\theta_l),f_p) &= \begin{bmatrix}
		&1 &\cdots
		&e^{-j\frac{2\pi\eta_p}{\lambda_c}\frac{(N-1)^2d^2}{2\mu(r_l,\theta_l)}}
	\end{bmatrix}^T
	\nonumber\\
	&= \begin{bmatrix}
		&1 &\cdots
		&e^{-j\frac{2\pi}{\lambda_c}\frac{(N-1)^2d^2}{2\mu(\tilde{r}_{l,p},\tilde{\theta}_{l,p})}}
	\end{bmatrix}^T
	\nonumber\\
	&= \boldsymbol{b}(\mu(\tilde{r}_{l,p},\tilde{\theta}_{l,p}),f_c),
	\label{str_decom2_2}
\end{align}
respectively, where
\begin{align}
	 \sin(\tilde{\theta}_{l,p}) &= \eta_p\sin(\theta_l),
	\nonumber\\
	 \tilde{r}_{l,p} &= \frac{1-\eta_p^2\sin^2(\theta_l)}
	{\eta_p\big(1-\sin^2(\theta_l)\big)},
	\nonumber\\
	 \mu(r_l,\theta_l) &= \eta_p\mu(\tilde{r}_{l,p},\theta_{l,p}),
	\label{spatial_loc}	
\end{align}
with $\eta_p \triangleq  {f_c}/{f_p}$. Substituting~\eqref{str_decom1_2} and~\eqref{str_decom2_2} into~\eqref{sparse_fc} will result in
\begin{align}
	\tilde{\boldsymbol{\beta}}_p &
	= \sum_{l=1}^{L}\tilde{g}_{l,p}\boldsymbol{D}_{\mu}^H\boldsymbol{a}(r_l,\theta_l,f_p)
	\notag\\
& = \sum_{l=1}^{L}\tilde{g}_{l,p} \boldsymbol{D}_{\mu}^H\left(\boldsymbol{b}(\mu(r_l,\theta_l),f_p) \circ \boldsymbol{a}(\theta_l,f_p) \right)   \notag\\
& =  \sum_{l=1}^{L}\tilde{g}_{l,p} \boldsymbol{D}_{\mu}^H \left(\boldsymbol{b}(\mu(\tilde{r}_{l,p},\tilde{\theta}_{l,p}),f_c) \circ \boldsymbol{a}(\tilde{\theta}_{l,p},f_c) \right)\notag\\
	&=
 \sum_{l=1}^{L}\tilde{g}_{l,p}\boldsymbol{D}_{\mu}^H \boldsymbol{\alpha} (\tilde{r}_{l,p},
	\tilde{\theta}_{l,p},f_c),
	\label{beta_split}
\end{align}

As shown in~\eqref{beta_split}, for the channel $\bm h_p$, the steering vector $\boldsymbol{a} (r_l,\theta_l,f_p)$ at the sub-carrier $f_p$ is equivalently transformed into a virtual steering vector $\boldsymbol{\alpha}(\tilde{r}_{l,p},\tilde{\theta}_{l,p},f_c)$ at the carrier frequency $f_c$. According to~\eqref{sparse_sc}, it can be derived that $\tilde{\boldsymbol{\beta}}$ also exhibits a block-sparse structure, with $\|\tilde{\boldsymbol{\beta}}_p \|_0$ being on the order of $1/\sqrt{N}$. Note that $\boldsymbol{\alpha}(\tilde{r}_{l,p}, \tilde{\theta}_{l,p},f_c)$ can be regarded as the steering vector for a virtual scatterer/user located at $\{\tilde{r}_{l,p},\tilde{\theta}_{l,p}\}$.
Consequently, the sparse structure of $\tilde{\boldsymbol{\beta}}_p$ is determined by these
virtual locations. At each sub-carrier, the virtual components split into different
spatial locations, leading to the frequency-variant sparse structures. Fortunately,
the intrinsic connection of these sparse structures can be found in
~\eqref{spatial_loc}.

It is clear that
spatial angle $\sin(\tilde{\theta}_{l,p})$ and the equivalent
effective distance $\mu(\tilde{r}_{l,p},\tilde{\theta}_{l,p})$
are both a liner function of the exact physical terms $\sin(\theta_l)$
and $\mu(r_l,\theta_{l})$ with the ratio $\eta_p$. Therefore,
the block-sparse pattern of the channels at different sub-carriers
change linearly against the frequency. This is to say, the
block-sparse pattern of adjacent sub-carriers are closely
coupled. By stacking the block-sparse representation of all
sub-carriers in a matrix, we can obtain a two-dimentional
block-sparse representation of the entire wideband channel
$\boldsymbol{H}$ on the modified DFT matrix $\boldsymbol{D}_{\mu}$,
i.e.,
\begin{align}
	\boldsymbol{H} = \boldsymbol{D}_{\mu} \tilde{\boldsymbol{B}},
	\label{sparse_all}
\end{align}
where $\tilde{\boldsymbol{B}}\triangleq \left[\tilde{\boldsymbol{\beta}}_1 \
\cdots \ \tilde{\boldsymbol{\beta}}_{P}\right] \in \mathbb{C}^{N_t \times P}$,
in which the non-zero entries are clustered in the dimensions of both
the support sets and the sub-carriers. We call this type of block-sparse
structure as 2D pattern-coupled block-sparsity.

\begin{figure}
	\centering
	\includegraphics[width=0.6\linewidth]{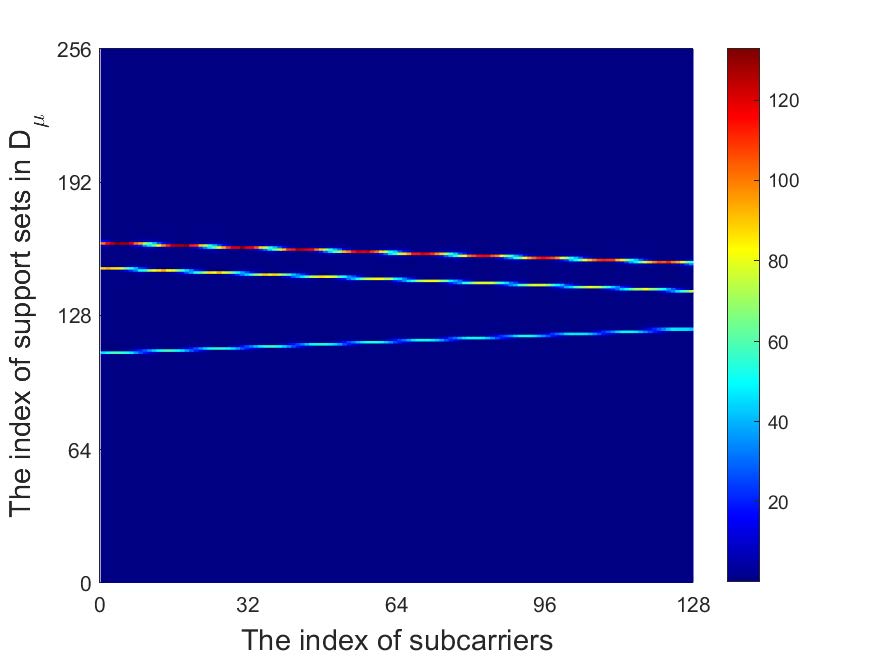}
	\caption{The sparsity of the wideband channel $\boldsymbol{H}$}
	\label{fig_sparsity}
\end{figure}

To verify the above sparsity properties, we plot in Fig. \ref{fig_sparsity}
the sparse structure of the wideband channel on a common modified
DFT dictionary. The simulation setup is set as following: $f_c = 100$ GHz,
$B = 10$ GHz, $P = 128$, $N = 256$ and $L = 3$. It can be seen that
each column consists of three non-zero blocks associated with the
three paths respectively. In addition, the blocks of different columns (
i.e., different sub-carriers) are closely coupled, showing a nearly linear
change.

\subsection{2D Block-Sparsity-Awared Wideband Near-field Channel Estimation}

As discussed in the former section, $\tilde{\boldsymbol{B}}$ has 2D block-sparse structure, and in this section, we adopt this specific sparse structure to devise an efficient channel estimation algorithm for a wideband THz XL-MIMO system. Substituting (\ref{sparse_all}) into
(\ref{ry_2}), we can obtain
\begin{align}
	\boldsymbol{Y} = \boldsymbol{F} \boldsymbol{D}_{\mu} \tilde{\boldsymbol{B}}
	+ \boldsymbol{N}
	= \boldsymbol{\Phi} \tilde{\boldsymbol{B}}
	+ \boldsymbol{N},
	\label{y_new}
\end{align}
where $\boldsymbol{\Phi}\triangleq \boldsymbol{F} \boldsymbol{D}_{\mu}$.
Therefore, the channel estimation problem can be converted to recover
the two-dimensional block-sparse matrix $\tilde{\boldsymbol{B}}$ with
the noisy observations $\boldsymbol{Y}$. To fully exploit the block-sparse
structure of each single-carrier channel as well as the correlation
between the channels of different sub-carriers, in this paper, we employ
the 2D pattern-coupled sparse Bayesian learning (PCSBL) algorithm
\cite{fang2016two} to recover the 2D block-sparse representation.


\section{Numerical Results}

In this section, we have conducted the numerical simulations to verify the
effectiveness of our proposed wideband channel estimation algorithm for a THz XL-MIMO system. In the simulations, $N$ is set to $256$; $f_c$ is set to $0.1$ THz; the bandwidth is $B = 10$ GHz; and the number of sub-carrier is set to $P = 128$. Each entry of the precoding matrix $\boldsymbol{F}$ are randomly generated from an independent and identically distributed (i.i.d.) complex Gaussian distribution, i.e., $f_{nt} \sim \mathbb{CN} (0,1/\sqrt{N})$. The signal-to-noise ratio (SNR) is defined as
$\text{SNR} \ (\text{dB}) = 10 \log_{10}\big({\mathbb{E}\{\|
\boldsymbol{H}\|_F^2\}}/{\mathbb{E}\{\|\boldsymbol{N}\|_F^2\}}\big)$. The number of paths is set to $L = 3$, which consists of a Line-of-sight (LoS) path and two non-LoS paths. The path gain follows the setting in \cite{jornet2011channel}.
The angle from the BS to the user and the distance at each path are uniformly distributed over $[-\pi/2,\pi/2]$ and $[10,20]$ m, respectively. The normalization mean square error (NMSE) is chosen as the performance metric, and it is defined as
\begin{align*}
\text{NMSE} \ (\text{dB})= 10\log_{10} \left(\mathbb{E}\left\{
 \frac{\|\hat{\boldsymbol{H}}-\boldsymbol{H}\|_F^2} {\|\boldsymbol{H}\|_F^2}
\right\}\right)
\end{align*}
where $\hat{\boldsymbol{H}}$ is the estimate of $\boldsymbol{H}$.

We compare the proposed 2D-PCSBL method with several state-of-art
methods, i.e., the LS algorithm, the near-field narrow-band
SOMP (NF-SOMP) algorithm~\cite{cui2022channel}, the bilinear pattern detection (BPD) based
near-field method (NF-BPD), and the subcarrier-to-subcarrier PCSBL algorithm (SC-PCSBL)~\cite{wang2023compressive}. Both the NF-SOMP and the NF-BPD are based on the polar-domain dictionary. However, the NF-BPD takes the beam-splitting effect into account, while the NF-SOMP does not. In the SC-PCSBL, the channels for different sub-carriers are estimated independently based on the block sparse structure.

\begin{figure}[!t]
	\centering
	\includegraphics[width=0.6\linewidth]{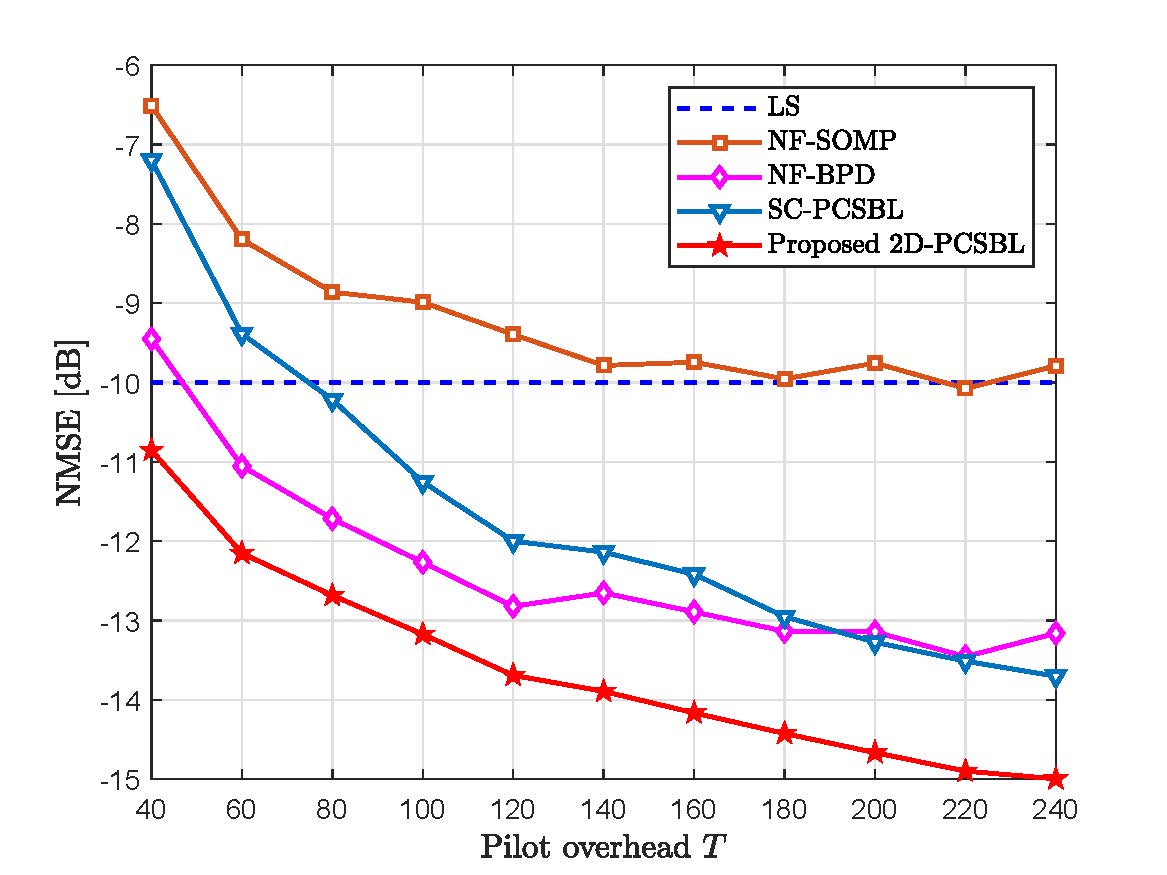}
	\caption{NMSE versus pilot overhead $T$ when SNR $= 10$ dB.}
	\label{fig_nmse_pilot}
\end{figure}

Fig.~\ref{fig_nmse_pilot} illustrates the NMSE performance of each algorithm against the length of the pilot sequence, where the SNR is fixed at $10$ dB. In the LS, the number of pilots is set sufficiently large ($T \gg NP$) to obtain satisfactory performance. It can be observed that the NMSE values of all these methods decrease as the pilot overhead increases. In addition, the NF-SOMP performs the worst because it assumes that different sub-carriers share the same sparse structure. The NF-BPD, which takes the beam-splitting effect into account, achieves a better NMSE than the SC-PCSBL. However, the SC-PCSBL shows a more significant improvement than the NF-BPD as the pilot overhead increases. When $T > 200$, the NMSE achieved by NF-BPD is no longer better than that of SC-PCSBL. Notably, the proposed 2D-PCSBL has the superiorityover the other methods, and the improvement comes from both the modified dictionary and the 2D-PCSBL algorithm.

\begin{figure}[!t]
	\centering
	\includegraphics[width=0.6\linewidth]{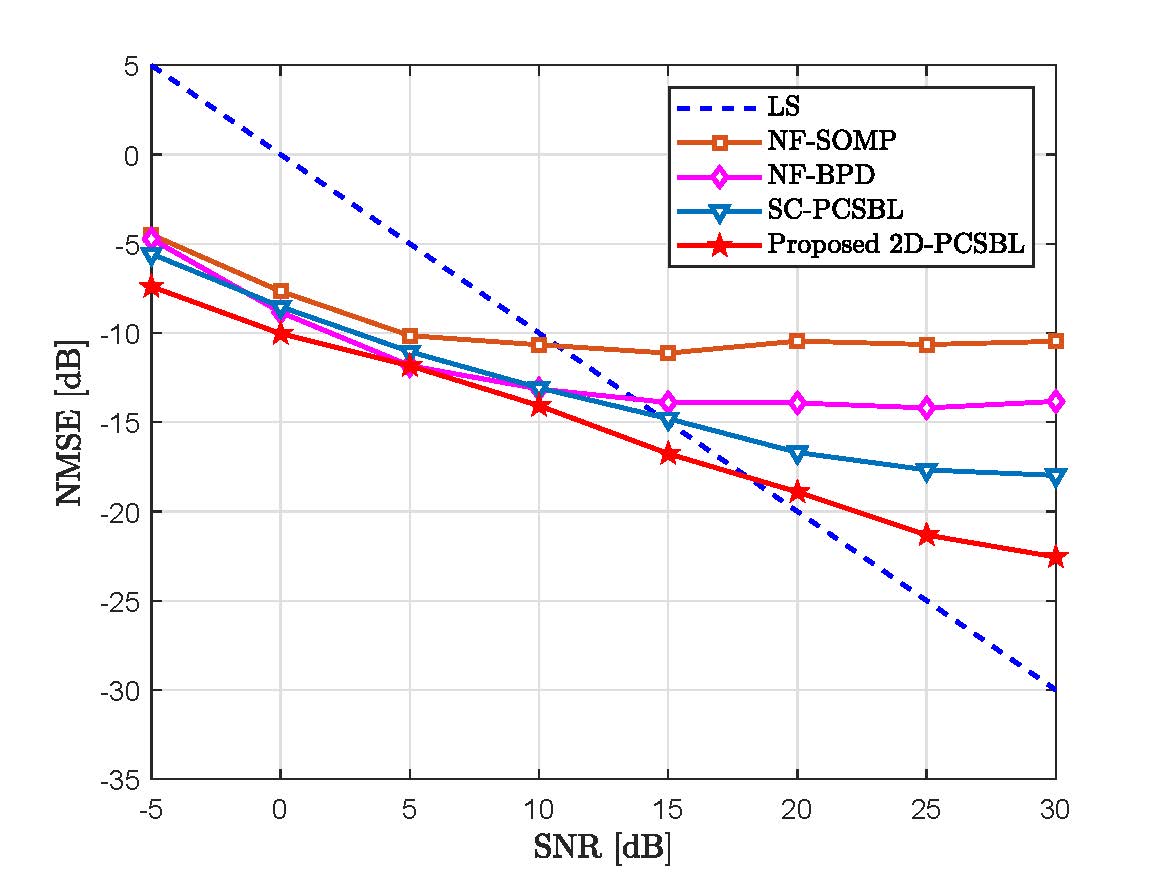}
    \caption{NMSE versus SNR when $T = 160$.}
	\label{fig_nmse_snr}
\end{figure}

The comparison of NMSE performances at different SNRs is depicted in Fig.~\ref{fig_nmse_snr}, where $T = 160$ is set identically for all the methods except for the LS algorithm. It can be observed that all the methods
show improved NMSE performances as the SNR increase. The NF-SOMP experiences a substantial performance loss since it ignores the beam-splitting effect. By leveraging the bilinear pattern of the beam-splitting effect, the NF-BPD achieves a better performance than the NF-SOMP. However, the NF-BPD performs significantly worse than the SC-PCSBL and the proposed 2D-PCSBL in the higher SNR region. In addition, the result also verifies the effectiveness of our proposed transform dictionary. When SNR is $30$dB, the gap between the proposed 2D-PCSBL and the NF-BPD is up to approximately $9$ dB. Furthermore, the proposed 2D-PCSBL is superior than the SC-PCSBL because the correlation between different sub-carriers is exploited to enhance the accuracy of wideband channel estimation.

In summary, the proposed 2D-PCSBL algorithm shows superiority over these comparative algorithms. The reasons can be articulated in two aspects. First of all, we construct a modified DFT matrix as our transform matrix, on which the wideband channel admits 2D pattern-coupled block-sparse structure. Compared with the polar-domain transform matrix, our proposed transform dictionary has the satisfactory RIP property with a moderate number of atoms. In addition, the 2D-PCSBL method can adaptively capture the connection between non-zero entries across different support sets and sub-carriers, which has the superiority than the other OMP-based methods.

\section{Conclusion}

The wideband near-field channel estimation problem for a THz XL-MIMO system was studied in this paper. We constructed a frequency-independent orthogonal dictionary such that the channel across sub-carriers was sparsely represented in a 2D block-sparse manner. As a result, we recast the considered channel estimation problem as a compressed sensing problem with a specific sparse structure. Numerical simulations were conducted to validate the performance of the proposed method. The results indicated its superiority over several existing near-field wideband estimation methods.

\bibliographystyle{IEEEtran}
\bibliography{IEEEabrv,newbib}
\end{document}